\documentclass[aps,prb,twocolumn,groupedaddress]{revtex4}

\usepackage{graphicx}

\newcommand{\celsius}{$^\circ$C}

\begin{document}

\title{First-principles calculation of the effect of strain on
the diffusion of Ge adatoms on Si and Ge (001) surfaces}

\author{A. van de Walle}
\email[]{avdw@alum.mit.edu}
\homepage[]{http://www.mit.edu/~avdw/}
\affiliation{Materials Science \& Engineering Department, Northwestern University, 
Evanston, IL 60208-3108}

\author{M. Asta}
\affiliation{Materials Science \& Engineering Department, Northwestern University, 
Evanston, IL 60208-3108}

\author{P. W. Voorhees}
\affiliation{Materials Science \& Engineering Department, Northwestern University, 
Evanston, IL 60208-3108}


\date{\today}

\begin{abstract}
First-principles calculations are used to calculate the strain 
dependencies of the binding and diffusion-activation energies for 
Ge adatoms on both Si(001) and Ge(001) surfaces. Our calculations 
reveal that the binding and activation energies on a strained Ge(001) 
surface increase and decrease, respectively, by 0.21 eV and 0.12 eV per 
percent compressive strain.  For a growth temperature of 600 \celsius, these 
strain-dependencies give rise to a 16-fold increase in adatom density 
and a 5-fold decrease in adatom diffusivity in the region of 
compressive strain surrounding a Ge island with a characteristic size 
of 10 nm.
\end{abstract}


\maketitle

In heteroepitaxial growth of lattice-mismatched thin films, the 
Stranski-Krastanov (SK) growth mode has been widely investigated
as a basis for self-assembling arrays of coherent nanostructured 
islands, commonly referred to as quantum dots (QDs).\cite{eaglesham:dfree}
The need for highly monodisperse QD 
arrays in semiconductor optoelectronic device applications has 
motivated extensive research into the microscopic mechanisms 
influencing the evolution of island size distributions during SK 
growth.  While the stabilizing thermodynamic effects associated 
with the elastic interactions between strained islands have been 
investigated in some detail (e.g., Refs. \onlinecite{bimberg,daruka}, and 
references cited therein), the role of various proposed 
{\em kinetic} mechanisms on the development of island size 
distributions remains less clear.

Within a self-consistent mean-field rate theory, Koduvely and 
Zangwill \cite{zangwill} demonstrated that with decreasing
island-island separation, a strain-mediated decrease in the barrier 
for adatom-island detachment leads to a reduction in the mean 
island size, with an associated narrowing of the size distribution.
These findings are qualitatively consistent with experimental 
observations \cite{Kobayashi} in InAs/GaAs.
Madhukar \cite{Madhukar} and Penev, Kratzer and Scheffler \cite{kratzer,penev:strain}
have considered the growth of InAs QDs on GaAs, where
increasing island size leads to a 
buildup of compressive elastic strains in the surrounding substrate.\cite{moll}
Within simplified models of 
diffusion-limited growth these authors demonstrated that,
in the InAs/GaAs heteroepitaxial system,
the strain-dependence of the parameters governing adatom diffusion 
gives rise to a reduction in the flux of adatoms reaching 
larger islands relative to small ones, leading to a reduced rate 
of coarsening and an associated narrowing of the size distribution.

While the potentially important consequences for island growth 
kinetics arising from strain dependencies in adatom binding and 
migration energies have been clearly demonstrated, attempts to 
determine the magnitude of these effects in specific systems 
have been undertaken in relatively few
systems.\cite{penev:strain,Ratsch,shu:simples,Roland,Spjut,Schroeder} 
These effects are most readily investigated using first-principles 
calculations \cite{kaxiras:revsurfd,smith:semiabi,
penev:strain,Ratsch,shu:simples}, as they are extremely difficult to 
isolate experimentally.  The purpose of the present work is to 
investigate the effect of strain-dependent adatom diffusion and 
binding energies upon island growth kinetics in Ge/Si(001).  This 
system represents one of the most widely studied examples of QD 
formation induced by SK growth, \cite{ross1,ross2,williams:review,floro,
Mo,tersoff:superlat}
yet to date the effect of strain upon Ge adatom binding energies 
and diffusion rates on Ge wetting layers remains unstudied.  We 
employ first-principles calculations to compute the strain 
dependence of Ge adatom energetics on Si(001) and Ge(001) 
surfaces.

\begin{figure}
\includegraphics{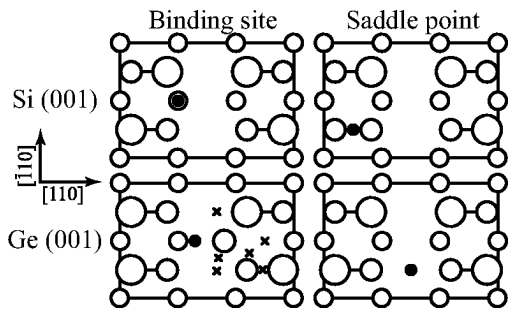}
\caption{\label{dimers}Geometries of the binding site and saddle 
point configurations for a Ge adatom (filled circle) diffusing on 
the Si (001) and Ge (001) surfaces.  Only the two topmost atomic
layers are shown. The size of each circle reflects the atom's
proximity to the observer. In the lower left 
quadrant, crosses indicate the adatom position for the 6 other 
candidate binding sites that were considered.}
\end{figure}

Our calculations were set up as follows.
We consider adatom diffusion in the dilute limit
(as opposed to dimer diffusion \cite{Zoethout}).
The initial atomic configuration for all of our calculations 
is taken to be the well-known minimum energy $c(4\times2)$
reconstruction
(depicted in the
upper left corner of Fig. \ref{dimers}).
The adatom calculations are performed using a supercell geometry where
an artificial three-dimensional periodicity is imposed on the
system. This artificial periodicity is harmless, provided that
convergence with respect to the distance between the periodic images
is achieved.  Following earlier studies of adatom diffusion of similar 
systems, \cite{Milman1,Milman2,Milman3} our supercell consists of
two repetitions of the surface unit cell, depicted in Fig. \ref{dimers}
along the dimer rows direction (i.e. a $4 \times 4$
supercell). Twelve layers of atoms were used, separated by 6 layers of
vacuum. By varying the
thickness of the slab and the width of the vacuum separating the periodic
images perpendicular to the surface, it was verified that the selected
system size provides energies with an accuracy of the order of 0.03 eV.
Single adatoms were
placed on each of the two slab surfaces, which has the effect
of doubling the adatom's energetic contribution, thereby improving
the accuracy of the calculations.  All calculations were performed 
using the {\it ab initio} total-energy and molecular-dynamics program VASP 
(Vienna ab initio simulation package) developed at the Institut 
f\"{u}r Materialphysik of the Universit\"{a}t Wien \cite{kresse:vasp1,
kresse:vasp2}, which implements Vanderbilt ultra-soft 
\cite{Vanderbilt:soft_pseudo} pseudo-potentials \cite{Phillips:pseudo}. 
The energy cutoff for the plane-wave basis set was set to 150 eV and a 
$2 \times 2 \times 1$ mesh of $k$-points was used for the $4 \times 4$ 
supercell. All atoms were allowed to relax.

On the Si $(001)$ surface, the diffusion of a Ge adatom along the
dimer rows is known to be of the order of 1000 times faster than
across dimer rows \cite{mo:sigeaniso} and we therefore focus solely 
on the diffusion along dimer rows. Since a typical surface consists 
of terraces where the direction of the dimer rows changes by 
90$^\circ$ at each monoatomic step, fast diffusion in any direction 
is possible at the mesoscopic level, even though the fast diffusion 
is unidirectional at the microscopic level. Earlier studies 
\cite{Milman1,Milman2,Milman3,dalphian:surfdof} have
unambiguously determined the location of the binding site and the
activated state of the Ge adatom on the Si $(001)$ surface (see Fig. 
\ref{dimers}) as well as the precise configuration of the surface 
dimers in the vicinity of the adatom.

\begin{figure}
\includegraphics{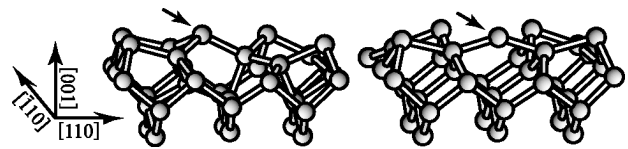}
\caption{\label{bndsadge}Geometries of the binding site (left) and saddle 
point (right) configurations for a Ge adatom (indicated by an arrow) diffusing on 
the Ge (001) surfaces.}
\end{figure}

Since experimental and computational evidence is scarcer in the case 
of the Ge adatom on the Ge $(001)$ surface, \cite{mae:sige} we considered 7 possible 
binding sites and found the minimum energy site to be as shown in Figures 
\ref{dimers} and \ref{bndsadge}. This site remains energetically favored
for values of the substrate lattice parameter up to 3\% larger than the Si
lattice parameter. Beyond that threshold, a binding site
analogous to the binding site of Ge on Si (001) becomes favorable.
We shall neglect this alternate binding site, as
the substrate lattice parameter required to stabilize it
falls outside the range sampled during Ge on Si (001) heteroepitaxy.

We identified the diffusion path using the nudged elastic band 
method \cite{mills:nudged,jonsson:nudged} and refined the position 
of the saddle point (see Figures \ref{dimers} and \ref{bndsadge}) using the quasi-Newton 
algorithm. \cite{kresse:vasp2}
We focused on the diffusion along the valleys between the dimer 
rows, because the binding sites located atop the dimer row had a 
binding energy at least 0.3 eV larger than the saddle point energy 
for diffusion within the valleys, strongly suggesting that any 
other saddle points located on the dimer rows would also have a 
higher energy than the saddle point we analyzed.
The nearly one-dimensional diffusion and the adatom's preference for
sites located in the valley between the dimer rows agrees qualitatively with the results of earlier 
calculations based on semi-empirical potentials, \cite{mae:sige}
although the precise location of the binding sites and of the saddle points differ.

\begin{figure}
\includegraphics{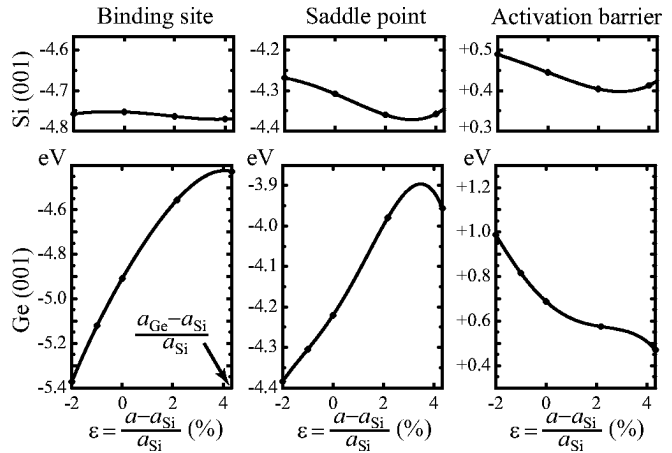}
\caption{\label{straindep}Strain-dependence of the binding energies
$E_b$ (left), the saddle point energies
$E_a+E_b$ (middle) and the activation barriers $E_a$ (right) for a Ge 
adatom diffusing on Si(001) (top) and Ge(001)
(bottom) surfaces. These plots are obtained by a 
polynomial fit to the calculated energies as a function strain.
Let $a$, $a_{\text{Si}}$ and $a_{\text{Ge}}$ respectively denote the substrate, the Si, and the Ge lattice parameters.}
\end{figure}

For each type of surface (Si or Ge), the energies of three geometries
were calculated (the free surface, the saddle point and the binding
site configurations) at various levels of biaxial strain imposed
parallel to the plane of the surface.
The resulting energy versus strain relationships are 
plotted in Fig. \ref{straindep}. Our results reveal four important observations.
First, linear approximations to the strain-dependence
of the binding energies and activation barriers \cite{shu:simples} must clearly be used with care.
Second, the Ge adatom binding energy does not necessarily exhibit a
minimum when the lattice parameter of the substrate matches the lattice parameter
of bulk Ge. 
Third, the sign of the change of these energies under strain are highly system-dependent:
calculations on the structurally similar In/GaAs(001) heteroepitaxial system \cite{penev:strain}
found the strain-dependences of the binding ($E_b$) and saddle-point energies ($E_b+E_a$)
the be of a sign opposite to the ones in the Ge/Ge(001) system.
The sign of the strain-dependence of the activation barrier ($E_a$) for Ge on both Si (001) and
Ge (001), however, agrees with earlier studies in In/GaAs(001) \cite{penev:strain} and Si/Si(001)
\cite{shu:simples,Roland,Spjut,Zoethout}.
Finally, the magnitude of the effect of strain on binding and saddle point energies found in the case
of Ge on Ge(001) is by far the largest,
relative to any other system studied so far.\cite{penev:strain,shu:simples,Ratsch,Roland,Spjut,Zoethout} 

\begin{figure}
\includegraphics{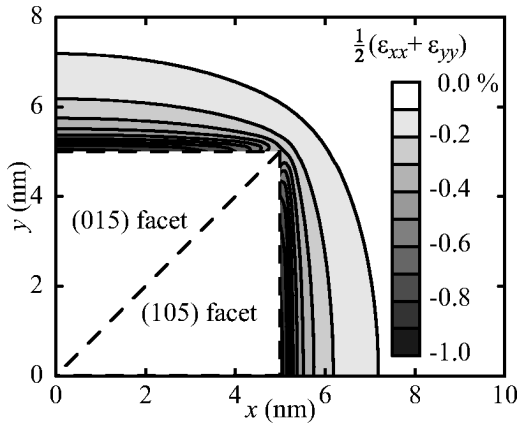}
\caption{\label{sfield}Strain field experienced by the substrate
surface in the vicinity of a Ge island (dotted lines) of a typical size (10 nm) 
terminated by $(105)$ facets. The homogenous component of the 
in-plane strain relative to the Si lattice parameter is plotted 
($(\epsilon_{xx}+\epsilon_{yy})/2$, for a surface normal to the $z$ axis).}
\end{figure}

In order to quantify the importance of these results in the context
of quantum dot growth, we calculated the strain field in the vicinity
of a Ge $(105)$-terminated pyramidal island with a characteristic size of 10 nm.\cite{williams}
For the calculation of strain fields around coherent
strained islands, both finite-element techniques (e.g., Ref. \onlinecite{moll}) as 
well as approximate analytical methods \cite{tersofftromp} have been
applied previously.  In the present work, elastic strain fields have
been derived from the results of the linear stability analysis of 
Spencer, Voorhees and Davis.\cite{spencer:linstab}  In this analysis, 
the linearized strain fields arising from Fourier-mode perturbations in 
the surface height were derived within isotropic elasticity theory.
By summing the resulting expressions over the Fourier amplitudes 
describing the shape function of a Ge pyramid, we calculated 
the in-plane strain field (shown in Fig. \ref{sfield})
at the surface of a Ge wetting layer
(three atomic-planes in thickness) on Si(001). In Fig.~\ref{sfield}
the strain is referenced to the Si lattice constant, so that zero
corresponds to a wetting layer epitaxially strained on the Si 
substrate.  In agreement with previous calculations for related
systems, \cite{moll,penev:strain,tersofftromp} the strain field 
surrounding the island is compressive in nature.

In Fig.~\ref{sfield}, the strain is seen to range in magnitude from 
zero to roughly $-1$\% in the vicinity of the Ge(105)-terminated island.  Over this 
range of compressive strains, the binding energy of a Ge adatom on a 
Ge surface varies by about 0.21 eV, while the activation barrier 
varies by about 0.12 eV, as seen in Fig. \ref{straindep}.  At a 
typical deposition temperature of 600 \celsius, these variations correspond, 
respectively, to a 16-fold variation in equilibrium adatom density and a 5-fold 
variation in adatom mobility. In contrast, the corresponding 
strain-dependences on the Si (001) surface are much less pronounced.
Of course, the strain-dependence of the entropic prefactors \cite{penev:strain} could very well
modify those order-of-magnitude estimates based solely on
the strain-dependence of the energetic contributions.

To further explore the implications of our calculated results, we
follow the analysis of Penev, Kratzer and Scheffler, \cite{penev:strain}
employing a simple surface diffusion-limited one-dimensional model of adatom diffusion between two 
neighboring islands.
We expand slightly on their work by including 
boundary conditions at the island edges that impose local equilibrium 
between adatoms and islands. For simplicity, we consider all entropic 
prefactors to be strain-independent. We perform a stability analysis by 
considering two islands of equal size located at $x=0$ and $x=l$ and calculate the changes in the 
adatom flux toward each island as the size of one is perturbed.
Let $F_{0},F_{l}$ denote the flux towards each island 
and consider the (dimensionless) asymmetry in the flux towards each 
island, $F=\left( F_{l}-F_{0}\right)/\left( 2\phi l \right)$, where 
$\phi $ is the adatom deposition rate and $l$ is the distance separating 
the two islands. Solving the above diffusion problem leads to 
\begin{equation}
F = -\frac{K}{M_{0}\phi l}\frac{\left( e^{\beta E_{i,l}}-e^{\beta
E_{i,0}}\right) }{l}-\frac{M_{1}}{M_{0}},  \label{eqF}
\end{equation}
where
\begin{equation}
M_{n} = \frac{1}{l^{n+1}}\int_{0}^{l}\left( x-\frac{l}{2}\right) ^{n}e^{\beta
\left( E_{a}\left( x\right) +E_{b}\left( x\right) \right) }dx.  \label{eqMn}
\end{equation}
and where $E_{a}\left( x\right)$ and $E_{b}\left( x\right)$ are,
respectively, the activation barrier and binding energy as a function
of position $x$; $E_{i,0}$ and $E_{i,l}$ are, respectively, the energies 
of an atom bound to the islands located at $x=0$ and $x=l$; $K$ is a 
constant incorporating all entropic prefactors and $\beta $ is reciprocal 
temperature $\left( k_{B}T\right) ^{-1}$.

Our analysis focuses on the diffusion of Ge adatoms on the Ge (001) 
surface, strained to match the lattice constant of the Si substrate,
since a Ge wetting layer covering the Si substrate is known to form in SK growth.
We assume that adatoms are insensitive to the presence of Si under the
Ge wetting layer, that no substantial Si-Ge interdiffusion occurs and that
surface reconstructions are unaffected by epitaxial strain. 
Accounting for the latter (e.g., through the inclusion of a strain-dependent
concentration of ``missing dimers''\cite{liu:sgemorph})
may provide another source of strain-dependent diffusion
deserving further consideration.

Under the above assumptions, the 
quantity $E_{a}\left( x\right) +E_{b}\left(x\right) $ decreases under 
a compressive strain, as shown in the middle panels of 
Fig.~\ref{straindep}.  When the two islands are identical, $F=0$, since 
$E_{i,l}=E_{i,0}$ and $ M_{1}=0 $ (as 
$E_{a}\left( x\right) +E_{b}\left( x\right) $ is then symmetric with 
respect to $x=\frac{l}{2}$).  Now, consider an increase in the size of 
the island at $x=l$.
The first term of Equation (\ref{eqF}) describes 
the thermodynamic driving force for coarsening: $e^{\beta E_{i,l}}$ decreases 
due to capillarity (a larger island has a smaller surface to volume 
ratio).  This term thus causes $F$ to become positive, increasing the 
flux towards the larger island.  This term is inversely proportional to 
$\phi$, demonstrating that increasing the deposition rate 
would act to reduce this natural coarsening effect.

The second term of Equation (\ref{eqF}), which is {\em independent}
of deposition flux, quantifies the effect of 
strain-dependent adatom binding and migration energies.
A slight increase in the size of the island at $x=l$ induces a compressive strain in the wetting layer
in the vicinity of that island, thus inducing a decrease in the saddle 
point energy $E_a(x)+E_b(x)$ in the same area and causing $M_1$ to become negative.
The result is a relative increase in the adatom flux 
towards the larger island.
In other words, this analysis suggests that the strain-dependencies
of Ge adatom binding and migration energies plotted in Fig.~\ref{straindep}
act to accelerate the rate of coarsening of larger islands.

In summary, first-principles calculations have been employed to compute
the strain dependence of Ge adatom binding energies and activation
energies for diffusion on both Si(001) and Ge(001) surfaces.  
For a growth temperature of 600 \celsius, these strain-dependencies give rise 
to a 16-fold increase in adatom density and a 5-fold decrease in adatom 
diffusivity in the region of compressive strain surrounding a Ge island 
with a characteristic size of 10 nm.  Within a simplified model of 
diffusion-limited growth, these strain dependencies are found to have the 
qualitative effect of accelerating the natural coarsening rate of larger 
islands, in contrast with earlier results obtained for the related InAs
on GaAs system.\cite{penev:strain} The large magnitude of the strain 
dependence of adatom energetics on Ge(001) obtained in the present calculations
also contrasts with earlier findings in other systems\cite{Ratsch,Roland,Spjut,penev:strain,shu:simples,Zoethout} and
warrants further consideration of these effects in more detailed kinetic models
to further elucidate their effect on island growth.  

\begin{acknowledgments}
This work was supported by the NSF under programs DMR-0102794 and NSF-MRSEC DMR-00706097, using computer
resources provided by the National Partnership for Advanced Computational 
Infrastructure at the University of Michigan.
\end{acknowledgments}


\end{document}